\documentclass[12pt,a4paper]{article}
\usepackage[latin1]{inputenc}
\usepackage{amsmath}
\usepackage{amsfonts}
\usepackage{amssymb}
\usepackage[normalem]{ulem}
\usepackage{hyperref}
\hypersetup{colorlinks=true}
\usepackage{cite}
\usepackage{geometry}
\begin{document}

\title{ Estimation of attenuation of gravitational waves by Bose-Einstein condensate dark matter halos using Bogoliubov-de Gennes theory}
\renewcommand{\thefootnote}{\fnsymbol{footnote}}
\author{Levent Akant\footnote{levent.akant@boun.edu.tr},
\.{I}. \c{C}a\u{g}r{\i} \.{I}\c{s}eri\footnote{islam.iseri@boun.edu.tr}
and \.{I}brahim Semiz\footnote{ibrahim.semiz@boun.edu.tr} \\
$\;$\\
\small Physics Department, Bo\u gazi\c ci University \\
\small 34342 Bebek, \.Istanbul, Turkey }
\date{}

\maketitle
\renewcommand{\thefootnote}{\arabic{footnote}}\setcounter{footnote}{0}

\begin{abstract}
We consider a gravitational plane wave passing through a galactic dark matter halo composed of weakly self-interacting, self-gravitating, Bose-Einstein condensate of ultralight particles. Treating the gravitational wave as a time dependent perturbation, we study energy transfer between the gravitational wave and the Bose-Einstein condensate by applying linear response theory to a non-uniform condensate described by the Bogoliubov-de Gennes theory, and compute the fractional loss in gravitational wave energy. We apply our results to investigate the extent to which this loss effects the estimation of the distance between the gravitational wave source and the earth. We show that the effect is negligible.
\end{abstract}

\section{Introduction}

Recently it was suggested~\cite{Fuent0, Fuent1, RS} that a desktop Bose-Einstein condensation (BEC) setup might be used to detect gravitational waves (GW), a possible alternative to interferometric detectors like LIGO. The GW would excite phonon modes in the BEC, which would be detected by standard methods. Of course this means that the BEC will absorb energy from the GW.

Another context in which BEC might be relevant is at a scale 20+ orders of magnitude larger; the dark matter problem. It is part of the standard lore in astrophysics and cosmology that galaxies are embedded in a halo of so-called dark matter (DM), i.e. matter that does not emit, absorb, scatter any measurable electromagnetic radiation. The total amount of dark matter is thought to be more than five times that of normal, i.e. baryonic matter, and its nature and composition is subject of ongoing investigation and speculation. One of the candidates considered is a BEC of ultralight ($\sim 10^{-22}\,\textrm{eV} - 1\, \textrm{eV}$ ) particles \cite{halo-of-DM1,halo-of-DM2,halo-of-DM3,BEC-rel-cont1,BEC-rel-cont2,ultralight_BEC_DM1,ultralight_BEC_DM2,ferreira,dwornik}. Since the BEC-based GW detector idea suggests that the BEC will absorb energy from a GW, it is natural to wonder how significant this absorbtion is when a plane GW passes through a galaxy embedded in such a DM halo; hence to wonder if this absorbtion might lead to measurable attenuation of the GW, leading to modification of GW distances. Motivated by this question in this paper we investigate energy transfer between a GW and a BEC of non-relativistic, self-gravitating, weakly self-interacting dark matter particles. Our strategy will be to employ linear response theory to compute the fractional energy loss of a GW passing through a galactic DM BEC described by the Bogoliubov-de Gennes (BdG) theory of nonuniform BEC.

A weak GW passing through BEC will act as a time dependent perturbation and drive the zero temperature BEC out of its ground state. Such a situation can be studied by linear response theory which takes into account the out of equilibrium statistical mechanics of the medium perturbed by a weak time dependent external field. Because of the gravitational self-interaction the BEC in the absence of the GW will not have a uniform condensate wavefunction. In this case BEC can be treated in BdG theory, as opposed to the Bogoliubov theory of uniform condensate. The condensate wavefunction is determined by the Gross-Pitaevskii equation. Following the literature \cite{wang, guzman0, chavanis, velten, boehmer,guzman} self-gravitation can be analyzed in the self-consistent mean field approximation and this leads to the Gross-Pitaevskii-Poisson (GPP) system which can be solved in the Thomas-Fermi (TF) approximation. On the other hand fluctuations around the TF solution of GPP system, which are needed in linear response theory, are determined by BdG equations.

In this work we will consider only dark matter with repulsive self-interactions. In general bosons with weak attractive self-interaction do not thermalize and do not form a stable condensate. In this case the use of the standard BdG theory is not appropriate for the analysis of the problem. However, in the case of self-gravitating dark matter axions with attractive self-interactions \cite{axion_as_DM2,axion_attractive,axiontherm, chakrabarty, axion_as_DM1}  there is theoretical evidence \cite{axion_attractive,axiontherm,chakrabarty} that they may thermalize and form a condensate as a result of their gravitational interactions (see however \cite{guth0}). Let us also note the following references on interactions of a GW with classical matter \cite{Khlopov, moretti1, moretti2}.

 In the next section we consider the Hamiltonian of a scalar field coupled to a weak external gravitational wave field and its non-relativistic limit. In Sec.3 we discuss the computation of energy dissipation using linear response theory. In Sec.4 we give a quick review of BdG theory. In Sec.5 we apply the BdG theory to a self-gravitating, weakly self-interacting Bose system. We study the GPP system, its solution in the TF approximation, the boundary layer structure, and the BdG equations. In Sec.6 we combine the linear response theory and BdG theory to derive our main results (\ref{fraction}) and (\ref{fracapprox}) for the fractional energy loss of a GW passing through a BEC medium. We apply our results to GW passing through a galactic BEC DM halo and obtain a numerical estimate of this fractional energy loss, leading to the conclusion given in the brief Sec.7. In the appendix we give a quick review of linear response theory.

\section{Bose-Einstein Condensate Perturbed by a Gravitational Wave}

Consider the gravitational wave in the Minkowski background
\begin{equation}
  g_{\mu\nu}=g^{(0)}_{\mu\nu}+h_{\mu\nu}.
\end{equation}
Here $g^{(0)}_{\mu\nu}=diag(-1,1,1,1)$ is the Minkowski metric with mostly positive signature and $h_{\mu\nu}$ is a small perturbation. We are going to take the direction of propagation of the gravitational wave to be the $z$-direction and work in the transverse traceless gauge:
\begin{equation}\label{metrich}
  h_{\mu\nu}=\left(
    \begin{array}{cccc}
      0 & 0 & 0 &0 \\
      0 & h(t-z/c) & 0 & 0 \\
      0 & 0 & -h(t-z/c) & 0 \\
      0 & 0 & 0 & 0 \\
    \end{array}
  \right)
\end{equation}
Of course, we have $h(t-z/c) << 1$. Consider now the action for the relativistic real scalar field with repulsive $\phi^{4}$ interaction,
\begin{displaymath}
  S = \frac{1}{c}\int d^{4}x \sqrt{-g}\,\left\{-\frac{1}{2}\,g^{\mu\nu}\partial_{\mu}\phi\partial_{\nu} \phi-\frac{m^{2}c^{2}}{\hbar^{2}}\phi^{2}-\frac{\lambda}{4!}\phi^{4}\right\}.
\end{displaymath}
Throwing away a vanishing boundary term,
\begin{equation}
 S = \frac{1}{c}\int d^{4}x \sqrt{-g}\,\left\{\frac{1}{2}\,\phi\left[g^{\mu\nu}\nabla_{\mu}\nabla_{\nu}-\frac{m^{2}c^{2}}{\hbar^{2}}\right]\phi-\frac{\lambda}{4!}\phi^{4}\right\}.
\end{equation}
We make in $S$ the substitution \cite{guth0,braaten,guth}
\begin{equation}
  \phi(x)=\sqrt{\frac{\hbar}{2m}}\left[e^{-i\frac{mc^{2}}{\hbar}t}\psi(x)+e^{i\frac{mc^{2}}{\hbar}t}\psi^{*}(x)\right],
\end{equation}
to take advantage of the fact that most of the time dependence is in the $e^{\pm i\frac{mc^{2}}{\hbar}t}$ terms in the nonrelativistic limit, i.e. $|\ddot{\psi}|<< mc^{2}|\dot{\psi}|/\hbar$. Then, the $e^{\pm 2i\frac{mc^{2}}{\hbar}t}$ terms oscillate very rapidly and therefore give no contribution to the integral (\textit{e.g.} by the Riemann-Lebesgue lemma). So, the action takes the nonrelativistic form
\begin{eqnarray}
  S_{NR} &=& \int dtd^{3}x\,\sqrt{\gamma}\,\left[\frac{i\hbar}{2}(\psi^{*}\dot{\psi}-\dot{\psi}^{*}\psi)+\frac{\hbar^{2}}{2m}\psi^{*}\Delta\psi-\frac{U_{0}}{2}(\psi^{*}\psi)^{2}\right].
\end{eqnarray}
Here $\gamma_{ij}=g_{ij}$ is the space-like part of the metric $g_{\mu\nu}$,
\begin{equation}
  \Delta =\gamma^{ij}\nabla_{i}\nabla_{j}=\frac{1}{\sqrt{\gamma}}\partial_{i}(\gamma^{ij}\sqrt{\gamma}\partial_{j}).
\end{equation}
is the Laplacian (for scalars) corresponding to $\gamma_{ij}$, and $U_{0}=\hbar^{4}\lambda/(8m^{2})$.

Thus in slightly perturbed Minkowski space the many body Hamiltonian is given as
\begin{equation}
  H_{\gamma}=\int\,d^{3}x\,\sqrt{\gamma}\left[-\frac{\hbar^{2}}{2m}\psi^{\dag}\Delta\psi+\frac{U_{0}}{2}|\psi|^{4}\right].
\end{equation}
Using the gravitational wave metric (\ref{metrich}) we have $\gamma_{ij}=g_{ij}=\delta_{ij}+h_{ij}$,
\begin{eqnarray}
  \sqrt{\gamma} &=& 1-\frac{h^{2}}{2}+O(h^{3}),
\end{eqnarray}
\begin{equation}
  \gamma^{ij}=\left(
    \begin{array}{ccc}
      1-h+h^{2}+O(h^{3}) & 0 & 0 \\
      0 & 1+h+h^{2}+O(h^{3}) & 0 \\
      0 & 0 & 1 \\
    \end{array}
  \right),
\end{equation}
and
\begin{equation}
  \Delta=\partial_{i}\partial_{i}+h(-\partial_{x}^{2}+\partial_{y}^{2})+h^{2}(\partial_{x}^{2}+\partial_{y}^{2})-hh'\partial_{z}+O(h^{3}).
\end{equation}
where $h'$ is the derivative of $h$ with respect to its argument. So in the presence of the gravitational wave the Hamiltonian can be written as
\begin{equation}
  H_{\gamma}=H+\int\,d^{3}x\sum_{a=1}^{3}J_{a}(\mathbf{x},t)\mathcal{O}_{a}(\mathbf{x})+O(h^{3}),
\end{equation}
with
\begin{equation} \label{hamilton12}
 H=\int d^{3}x\,\mathcal{H},\;\;\;\;\;\;\mathcal{H}=-\frac{\hbar^{2}}{2m}\psi^{\dag}\delta^{ij}\partial_{i}\partial_{j}\psi+\frac{u_{0}}{2}|\psi|^{4},
\end{equation}
and
\begin{eqnarray}
  J_{1}(t,\mathbf{x}) &=& h\left(t-\frac{z}{c}\right),\;\;J_{2}(t,\mathbf{x})=h^{2}\left(t-\frac{z}{c}\right),\;\;J_{3}(t,\mathbf{x})=-h\left(t-\frac{z}{c}\right)h'\left(t-\frac{z}{c}\right), \\
  \mathcal{O}_{1}(\mathbf{x}) &=&\psi^{\dag}(\mathbf{x})\left[-\frac{\hbar^{2}}{2m}(-\partial_{x}^{2}+\partial_{y}^{2})\right]\psi(\mathbf{x}) \label{po1}  \\
  \mathcal{O}_{2}(\mathbf{x}) &=& -\frac{1}{2}\mathcal{H}+\psi^{\dag}(\mathbf{x})\left[-\frac{\hbar^{2}}{2m}(\partial_{x}^{2}+\partial_{y}^{2})\right]\psi(\mathbf{x})\\
  \mathcal{O}_{3}(\mathbf{x}) &=& \psi^{\dag}(\mathbf{x})\left[-\frac{\hbar^{2}}{
  2m}\partial_{z}\right]\psi(\mathbf{x}).
\end{eqnarray}
Note that $J_{2}$ and $J_{3}$ are of quadratic order in $h$. Also we will take $h(t-z/c)\rightarrow 0$ as $|t|\rightarrow\infty$ which implies $J_{i}(t,\mathbf{x})\rightarrow 0$ as $|t|\rightarrow\infty$.

\section{Energy Dissipation in Linear Response Theory}
We will compute the energy dissipated in the Bose field by using linear response theory, appropriate since the gravitational wave is a time dependent perturbation acting on the Bose-Einstein condensate which is assumed to be in the vacuum state $|B\rangle$ in remote past before the gravitational wave arrives. A brief review of the general method is given in the Appendix. In linear response theory the dissipated energy $\Delta E$ up to second order in $h$ is given as
\begin{eqnarray}
  \Delta E &=& \sum_{i=1}^{3}\int dt\,\dot{J}_{i}(t,\mathbf{x})\langle B|\mathcal{O}_{i}(\mathbf{x})|B\rangle\nonumber\\&& -\int dt\,d^{3}x\, dt'\,d^{3}x'\,\dot{J}_{1}(t,\mathbf{x})J_{1}(t',\mathbf{x'})\chi(t-t',\mathbf{x},\mathbf{x'}),
\end{eqnarray}
where $\dot{J}_{i}(t,\mathbf{x})=\partial_{t}J_{i}(t,\mathbf{x})$,
\begin{equation}\label{response}
  \chi(t-t',\mathbf{x},\mathbf{x'})=\frac{i}{\hbar}\theta(t-t')\langle B|[\mathcal{O}_{1H}(\mathbf{x},t),\mathcal{O}_{1H}(\mathbf{x'},t')]|B\rangle
\end{equation}
is the linear response function, and
\begin{equation}\label{Oheisenberg}
  \mathcal{O}_{1H}(\mathbf{x},t)=e^{\frac{i}{\hbar}Ht}\mathcal{O}_{1}(\mathbf{x})e^{-\frac{i}{\hbar}Ht}
\end{equation}
 is the Heisenberg picture operator defined in reference to the unperturbed (that is, unperturbed by the GW) Hamiltonian $H$ given in (\ref{hamilton12}). But since $J_{i}(t,\mathbf{x})\rightarrow 0$ as $|t|\rightarrow\infty$ we get
\begin{eqnarray}
  \Delta E = -\int dt\,d^{3}x\, dt'\,d^{3}x'\,\dot{h}\left(t-\frac{z}{c}\right)h\left(t'-\frac{z'}{c}\right)\chi(t-t',\mathbf{x},\mathbf{x'}),
\end{eqnarray}
We also define
\begin{equation}
  \chi(\omega,\mathbf{x},\mathbf{x'})=\lim_{a\rightarrow 0^{+}}\int dt\,e^{i\omega t}e^{-a t}
  \chi(t,\mathbf{x},\mathbf{x'}),
\end{equation}
and
\begin{equation}
  \widetilde{h}(\omega)=\int dt\,e^{i\omega t}h(t).
\end{equation}
Then in the frequency domain we have
\begin{equation}
  \Delta E=-i\int \frac{d\omega}{2\pi}\,\omega\,|h(\omega)|^{2}\int d^{3}x\,d^{3}x'\,e^{-i\frac{\omega}{c}(z-z')}\chi(\omega,\mathbf{x},\mathbf{x'}).
\end{equation}
Note that $\omega|h(\omega)|^{2}=\omega h^{*}(\omega)h(\omega)=\omega h(-\omega)h(\omega)$ is an odd function of $\omega$. So we can write $\Delta E$ as
\begin{equation}\label{ediss}
  \Delta E=-i\int \frac{d\omega}{2\pi}\,\omega\,|h(\omega)|^{2}\left[\int d^{3}x\,d^{3}x'\,e^{-i\frac{\omega}{c}(z-z')}\chi(\omega,\mathbf{x},\mathbf{x'})\right]_{\textrm{odd}},
\end{equation}
where the subscript odd means the odd part of the function of $\omega$ appearing inside the square brackets.

We will investigate $\Delta E$ in Bogoliubov-de Gennes (BdG) theory, where the unperturbed Hamiltonian $H$ will be approximated by an appropriate quadratic Hamiltonian.

\section{Bogoliubov-de Gennes Theory}\label{BdGSection}
Consider the many-body Hamiltonian
\begin{eqnarray}
  H &=& \int d^{3}x\,\left[\psi^{\dag}(\mathbf{x})\widehat{h}\psi(\mathbf{x})+\frac{U_{0}}{2} |\psi^{\dag}(\mathbf{x})\psi(\mathbf{x})|^{2}\right],
\end{eqnarray}
where
\begin{equation}
  \widehat{h}=-\frac{\hbar^{2}}{2m}\nabla^{2}+V_{\textrm{ext}}-\mu.
\end{equation}
Here $V_{\textrm{ext}}$ is an external potential. The first step of the BdG theory \cite{pethick} is the expansion of the field operator $\psi$ around a $c$-number background field $\phi$, which represents the condensate wavefunction,
\begin{equation}\label{fieldexp}
  \psi(\mathbf{x})=\phi(\mathbf{x})+\eta(\mathbf{x}),
\end{equation}
where $\eta$ denotes the quantum fluctuations around $\phi$. The condensate wavefunction $\phi$ is normalized as
\begin{equation}\label{normalizationphi}
  N_{0}=\int d^{3}x\,|\phi(\mathbf{x})|^{2},
\end{equation}
where $N_{0}$ is the number of condensed particles. Since $\phi$ is a $c$-number field $\eta$ and $\eta^{\dag}$ satisfy the canonical commutation relations
\begin{eqnarray}\label{ccr}
  [\eta(\mathbf{x}),\eta^{\dag}(\mathbf{x}')] = \delta(\mathbf{x}-\mathbf{x}'),\;\;\;\;\;[\eta(\mathbf{x}),\eta(\mathbf{x}')]=0=[\eta^{\dag}(\mathbf{x}),\eta^{\dag}(\mathbf{x}')].
\end{eqnarray}
Thus $H$ takes the form,
\begin{equation}
  H=H^{(0)}+H^{(1)}+H^{(2)}+H^{(3)}+H^{(4)},
\end{equation}
where $H^{(k)}$ is of order $k$ in the field operators $\eta$ and $\eta^{\dag}$. In BdG theory we ignore cubic and quartic terms $H^{(3)}$ and $H^{(4)}$ and approximate $H$ as
\begin{eqnarray}
  H &=& H^{(0)}+H^{(1)}+H^{(2)}.
\end{eqnarray}
The explicit forms of the individual terms appearing in the above expression are
\begin{eqnarray}
  H^{(0)} &=& \int d^{3}x\,\left\{\phi^{*}(\mathbf{x})\widehat{h}\phi(\mathbf{x})+\frac{U_{0}}{2} |\phi^{*}(\mathbf{x})\phi(\mathbf{x})|^{2}\right\},\label{line1}\\
  H^{(1)} &=& \int d^{3}x\,\left\{\phi^{*}(\mathbf{x})\widehat{h}\eta(\mathbf{x})+\eta^{\dag}(\mathbf{x})\widehat{h}\phi(\mathbf{x})+U_{0}|\phi(\mathbf{x})|^{2}\left[\eta^{\dag}(\mathbf{x})\phi(\mathbf{x})+
  \eta(\mathbf{x})\phi^{*}(\mathbf{x})\right]\right\},\label{line2}\\
  H^{(2)} &=& \int d^{3}x\,\left\{\eta^{\dag}(\mathbf{x})\widehat{h}\eta(\mathbf{x})+\frac{U_{0}}{2} \left[4|\phi(\mathbf{x})|^{2}\eta^{\dag}(\mathbf{x})\eta(\mathbf{x})+
  \eta^{\dag\,2}\phi^{2}(\mathbf{x})+\eta^{2}\phi^{*\,2}(\mathbf{x})\right]\right\}.\label{line3}
\end{eqnarray}
Now the background $\phi$ is chosen so as to make $H^{(1)}$ vanish. This gives us the Gross-Pitaevskii equation for the condensate wavefunction $\phi$
\begin{equation}\label{GP}
  \left[-\frac{\hbar^{2}}{2m}\nabla^{2}+V_{\textrm{ext}}-\mu\right]\phi(\mathbf{x})+U_{0}|\phi(\mathbf{x})|^{2}\phi(\mathbf{x})=0.
\end{equation}
Thus we arrive at the BdG Hamiltonian,
\begin{equation}\label{HBDG}
  H_{\textrm{BdG}}=H^{(0)}+H^{(2)}.
\end{equation}
The quadratic part $H^{(2)}$ of $H_{\textrm{BdG}}$, which contains the terms $\eta^{2}$, $\eta^{\dag\,2}$, is diagonalized by a generalized Bogoliubov transformation
\begin{equation}\label{etas0}
  \eta(\mathbf{x})=\sum_{r}u_{r}(\mathbf{x})b_{r}+v_{r}^{*}(\mathbf{x})b^{\dag}_{r},\;\,\;\;\;\; \eta^{\dag}(\mathbf{x})=\sum_{r}u_{r}^{*}(\mathbf{x})b^{\dag}_{r}+v_{r}(\mathbf{x})b_{r}.
\end{equation}
Here $b_{r}$ and $b_{r}^{\dag}$ are bosonic creation and annihilation operators $[b_{r},b_{s}^{\dag}]=\delta_{rs}$, $[b_{r},b_{s}]=0=[b^{\dag}_{r},b_{s}^{\dag}]$. Using (\ref{etas0}) in (\ref{ccr}) we get the following relations among the mode functions $u_{r}$ and $v_{r}$
\begin{eqnarray}
  \sum_{r} u_{r}^{*}(\mathbf{x}')u_{r}(\mathbf{x})-v_{r}^{*}(\mathbf{x}')v_{r}(\mathbf{x}) &=& \delta(\mathbf{x}-\mathbf{x}') \\
  \sum_{r} v_{r}^{*}(\mathbf{x}')u_{r}(\mathbf{x})-v_{r}^{*}(\mathbf{x})u_{r}(\mathbf{x}')  &=& 0.
\end{eqnarray}
On the other hand using (\ref{etas0}) in (\ref{H2}) and requiring $H^{(2)}$ to be in the form of a Hamiltonian of a system of decoupled harmonic oscillators
\begin{equation}
  H^{(2)}=\sum_{r}\epsilon_{r}(b^{\dag}_{r}b_{r}+b_{r}b^{\dag}_{r}).
\end{equation}
 we arrive at the BdG equations for the mode functions $u_{r}$, $v_{r}$ and the energy eigenvalues $\epsilon_{r}$,
\begin{eqnarray}\label{BdG}
  (\widehat{h}+2U_{0}|\phi|^{2})u_{r}+U_{0}\phi^{2}v_{r} &=& \epsilon_{r}u_{r} \\
   U_{0}\phi^{*\,2}u_{r}+ (\widehat{h}+2U_{0}|\phi|^{2})v_{r}&=&-\epsilon_{r}v_{r}.
\end{eqnarray}
The ground state of the $H_{BdG}$ is therefore the Bogoliubov vacuum $|B\rangle$ which is annihilated by all $b_{r}$'s
\begin{equation}
  b_{r}|B\rangle=0,\;\;\;\;\textrm{for all}\;\;r.
\end{equation}
In the Heisenberg picture defined by $H_{\textrm{BdG}}$ the fluctuation field is given in terms of the mode functions and the corresponding eigenvalues as
\begin{eqnarray}\label{etahe}
  \eta(\mathbf{x},t)= \sum_{r} u_{r}(\mathbf{x})e^{-i\omega_{r}t}b_{r}+v_{r}^{*}(\mathbf{x})e^{i\omega_{r}t}b_{r}^{\dag},\;\;\;\;
  \eta^{\dag}(\mathbf{x},t)= \sum_{r} u^{*}_{r}(\mathbf{x})e^{i\omega_{r}t}b^{\dag}_{r}+v_{r}(\mathbf{x})e^{-i\omega_{r}t}b_{r}.
\end{eqnarray}
Also note that the expectation value of the number density operator $\langle\psi^{\dag}(\mathbf{x})\psi(\mathbf{x})\rangle$ in the Bogoliubov vacuum $|B\rangle$
is given by
\begin{equation}\label{numden}
  \langle \psi^{\dag}(\mathbf{x})\psi(\mathbf{x})\rangle=|\phi(\mathbf{x})|^{2}+\langle \eta^{\dag}(\mathbf{x})\eta(\mathbf{x})\rangle.
\end{equation}

As a special case let us consider the Bogoliubov theory of a uniform condensate. In this case the condensate wavefunction and the chemical potential are given as
\begin{equation}
  \phi(\mathbf{x})=\sqrt{n_{0}},\;\;\;\;\;\mu=U_{0}n_{0},
\end{equation}
where $n_{0}$ is the number density of condensed particles, and (\ref{BdG}) can be solved exactly
\begin{equation}
  u_{r}(\mathbf{x})=\cosh\theta_{\mathbf{k}}e^{-i\mathbf{k\cdot\mathbf{x}}},\;\;\;\;\; v_{r}(\mathbf{x})=\sinh\theta_{\mathbf{k}}e^{-i\mathbf{k\cdot\mathbf{x}}},
\end{equation}
with
\begin{eqnarray}\label{coeff}
  \cosh\theta_{\mathbf{k}} = \sqrt{\frac{\epsilon^{0}_{\mathbf{k}}+U_{0}n_{0}}{2\epsilon_{\mathbf{k}}}+\frac{1}{2}},\;\;\;\;\;\; \sinh\theta_{\mathbf{k}}= \sqrt{\frac{\epsilon^{0}_{\mathbf{k}}+U_{0}n_{0}}{2\epsilon_{\mathbf{k}}}-\frac{1}{2}},
\end{eqnarray}
and
\begin{equation}\label{disp}
  \epsilon_{\mathbf{k}}=\sqrt{(\epsilon^{0}_{\mathbf{k}})^{2}+2 U_{0}n_{0}\epsilon^{0}_{\mathbf{k}}},\;\;\;\;\;\;\epsilon^{0}_{\mathbf{k}}=\frac{\hbar^{2}k^{2}}{2m}.
\end{equation}

For large chemical potential (or equivalently for low momenta)  $\epsilon^{0}(k)<<\mu=U_{0}n_{0}$  the dispersion relation (\ref{disp}) is approximated by phonon-like dispersion relation
\begin{equation}
  \epsilon_{\mathbf{k}}\simeq \sqrt{\frac{U_{0}n_{0}}{m}}\hbar k.
\end{equation}

On the other hand for small chemical potential (or equivalently for high momenta) $\epsilon^{0}(k)>>\mu=U_{0}n_{0}$  we have
\begin{equation}\label{dispsmallmu}
  \epsilon_{\mathbf{k}}\simeq \epsilon^{0}_{\mathbf{k}}=\frac{\hbar^{2}k^{2}}{2m},
\end{equation}
and
\begin{equation}
  \cosh\theta_{\mathbf{k}}\simeq 1,\;\;\;\;\;\;\sinh\theta_{\mathbf{k}}\simeq 0,
\end{equation}
\begin{equation}
  u_{r}(\mathbf{x})\simeq e^{-i\mathbf{k\cdot\mathbf{x}}},\;\;\;\;\; v_{r}(\mathbf{x})\simeq 0.
\end{equation}
Obviously the approximate results for the case $\epsilon^{0}(k)>>\mu=U_{0}n_{0}$  can also be obtained as the leading order perturbative solution of the BdG equations (\ref{BdG}) in the parameter $U_{0}n_{0}$.

\section{Self-Gravitation}
For a Bose system which self-interacts not only through a hard core potential but also by gravity we have
\begin{equation}
  V(\mathbf{x}-\mathbf{x}')=U_{0}\delta(\mathbf{x}-\mathbf{x}')+V_{\textrm{g}}(\mathbf{x}-\mathbf{x}'),
\end{equation}
with
\begin{equation}
  V_{\textrm{g}}(\mathbf{x}-\mathbf{x}')=-\frac{Gm^{2}}{|\mathbf{x}-\mathbf{x}'|}.
\end{equation}
Thus the many-body Hamiltonian is given as
\begin{eqnarray}
  H &=& \int d^{3}x\,\left[\psi^{\dag}(\mathbf{x})\left(-\frac{\hbar^{2}}{2m}\nabla^{2}-\mu\right)\psi(\mathbf{x})+\frac{U_{0}}{2}|\psi(\mathbf{x})|^{4}\right]+H_{\textrm{g}},
\end{eqnarray}
where
\begin{equation}
  H_{\textrm{g}}=\frac{1}{2}\int d^{3}x\int d^{3}x'\,
  \psi^{\dag}(\mathbf{x})\psi^{\dag}(\mathbf{x}')V_{\textrm{g}}(\mathbf{x}-\mathbf{x}')\psi(\mathbf{x}')\psi(\mathbf{x}).
\end{equation}
Now we apply Hartree approximation to the self-gravitation term $H_{\textrm{g}}$ and replace it by the mean field Hamiltonian
\begin{eqnarray}
H_{\textrm{gmf}}= \frac{1}{2}\int d^{3}x\int d^{3}x'\,
  \psi^{\dag}(\mathbf{x})V_{\textrm{g}}(\mathbf{x}-\mathbf{x}')\langle\psi^{\dag}(\mathbf{x}')\psi(\mathbf{x}')\rangle\psi(\mathbf{x}).
  \end{eqnarray}
Note that we are applying the Hartree approximation only to the self-gravitating part of $H$ and not to the hard-core self-interaction part, instead we will analyze the latter using the BdG theory. Thus we apply BdG theory to the resulting many-body Hamiltonian. Using (\ref{fieldexp}) and (\ref{numden}) we can write $H_{\textrm{gmf}}$ as
\begin{equation}
 H_{\textrm{gmf}}= \frac{1}{2}\int d^{3}x
  \psi^{\dag}(\mathbf{x})\left\{\int d^{3}x'\,V_{\textrm{g}}(\mathbf{x}-\mathbf{x}')
  \left[|\phi(\mathbf{x}')|^{2}+\langle\eta^{\dag}(\mathbf{x}')\eta(\mathbf{x}')\rangle\right]\right\}\psi(\mathbf{x}).
\end{equation}
Neglecting third and fourth order terms in fluctuations according to the general prescription of BdG theory, we arrive at the approximation
\begin{equation}
H\simeq H^{(0)}+H^{(1)}+H^{(2)}
\end{equation}
where
\begin{eqnarray}
  H^{(0)} &=& \int d^{3}x\,\left\{\phi^{*}(\mathbf{x})\left[\widehat{h}_{1}+\int d^{3}x'\,V_{g}(\mathbf{x}-\mathbf{x}')
  \langle\eta^{\dag}(\mathbf{x}')\eta(\mathbf{x}')\rangle\right]\phi(\mathbf{x})+\frac{U_{0}}{2}|\phi(\mathbf{x})|^{4}\right\}, \\
  H^{(1)} &=& \int d^{3}x\,\left\{\phi^{*}(\mathbf{x})\widehat{h}_{1}\eta(\mathbf{x})+\eta^{\dag}(\mathbf{x})\widehat{h}_{1}\phi(\mathbf{x})+U_{0}|\phi(\mathbf{x})|^{2}\left[\eta^{\dag}(\mathbf{x})\phi(\mathbf{x})+
  \eta(\mathbf{x})\phi^{*}(\mathbf{x})\right]\right\},\\
  H^{(2)} &=& \int d^{3}x\,\left\{\eta^{\dag}(\mathbf{x})\widehat{h}_{1}\eta(\mathbf{x})+\frac{U_{0}}{2} \left[4|\phi(\mathbf{x})|^{2}\eta^{\dag}(\mathbf{x})\eta(\mathbf{x})+
  \eta^{\dag\,2}\phi^{2}(\mathbf{x})+\eta^{2}\phi^{*\,2}(\mathbf{x})\right]\right\},\label{H2}
\end{eqnarray}
and
\begin{equation}
  \widehat{h}_{1}=-\frac{\hbar^{2}}{2m}\nabla^{2}-\mu+V_{\textrm{sc}}(\mathbf{x}),\;\;\;\;\;\;\;V_{\textrm{sc}}(\mathbf{x})=\int d^{3}x'\,V_{\textrm{g}}(\mathbf{x}-\mathbf{x}')|\phi(\mathbf{x}')|^{2}.
\end{equation}
Now the condition $H^{(1)}=0$ leads to the self-consistent Gross-Pitaevskii-Poisson (GPP) system
\begin{eqnarray}
  \left[-\frac{\hbar^{2}}{2m}\nabla^{2}-\mu+V_{\textrm{sc}}(\mathbf{x})+U_{0}|\phi(\mathbf{x})|^{2}\right]\phi(\mathbf{x}) = 0 \label{GPP1}\\
  V_{\textrm{sc}}(\mathbf{x}) = -Gm^{2} \int d^{3}x'\,\frac{|\phi(\mathbf{x}')|^{2}}{|\mathbf{x}-\mathbf{x}'|}\label{GPP2},
\end{eqnarray}
which will be treated in the Thomas-Fermi approximation in the next section.

Comparing the results of this section with (\ref{line1}), (\ref{line2}), (\ref{line3}) we see that, apart from the $V_{\textrm{g}}$ term in $H^{(0)}$, $H^{(0)}+H^{(2)}$ is the Hamiltonian of the BdG theory in an external potential $V_{\textrm{sc}}(\mathbf{x})$ which is determined in a self-consistent manner by the condensate wave-function. On the other hand $H^{(0)}$ effects only the ground state energy of the system, which will not play any role in our considerations, and therefore can simply be ignored in what follows. Thus we arrive at
\begin{equation}
  H_{\textrm{BdG}}=H^{(2)}.
\end{equation}

\subsection{Thomas-Fermi Approximation}
The GPP system is well studied in the literature using the Thomas-Fermi approximation \cite{wang, guzman0, chavanis, velten, boehmer,guzman} which is based on the assumption that the kinetic term in (\ref{GPP1}) is negligible,
\begin{equation}\label{ThomasFermi}
  -\mu+V_{sc}(\mathbf{x})+U_{0}|\phi(\mathbf{x})|^{2}=0.
\end{equation}
Now taking the Laplacian of this equation and using (\ref{GPP2}) we find
\begin{equation}
  \nabla^{2}|\phi(\mathbf{x})|^{2}=\frac{4\pi Gm^{2}}{U_{0}}|\phi(\mathbf{x})|^{2}.
\end{equation}
The spherically symmetric real solution to this equation is
\begin{equation}\label{GPSOL}
  \phi_{TF}(r)=\sqrt{C_{0}\,\frac{\sin k_{0}r}{k_{0}r}},
\end{equation}
where
\begin{equation}
  k_{0}=\sqrt{\frac{Gm^{3}}{\hbar^{2}a}}.
\end{equation}
The condensate wavefunction is then given as
\begin{equation}
  \phi(\mathbf{x})=\phi_{TF}(r)\Theta(R_{0}-r),
\end{equation}
where $R_{0}$ is the radius  of the dark matter halo given by the condition $\phi_{TF}(R_{0})=0$
\begin{equation}\label{R0}
  R_{0}=\frac{\pi}{k_{0}}=\pi\sqrt{\frac{U_{0}}{4\pi Gm^{2}}}.
\end{equation}
The constant $C_{0}$ is determined by the normalization condition (\ref{normalizationphi}) as
\begin{equation}\label{C0}
  C_{0}=\frac{N_{0}k_{0}^{3}}{4\pi^{2}}.
\end{equation}
The chemical potential in TF approximation can be calculated by evaluating (\ref{ThomasFermi}) at $\mathbf{x}=0$,
\begin{equation}\label{mutf}
  \mu_{TF}=-U_{0}C_{0}=-N_{0}\frac{Gm^{2}}{R_{0}}=-\frac{GMm}{R_{0}},
\end{equation}
where in writing the second equality we used (\ref{R0}) and (\ref{C0}), and in the last equality we defined $M=N_{0}m$ as the total mass of the condensate.

\subsection{Bogoliubov-de Gennes Equations}
The quadratic $H^{(2)}$ term given in (\ref{H2}) can be put in the standard form that does not contain squares of fluctuating fields, $\eta^{2}$ and $(\eta^{\dag})^{2}$, exactly as in the case of a uniform condensate by using the Bogoliubov transfomation (\ref{etas0}). In this case the BdG equations are given as
\begin{eqnarray}\label{BdGTF}
  \left[-\frac{\hbar^{2}}{2m}\nabla^{2}-\mu+V_{ext}(\mathbf{x})+2U_{0}|\phi(\mathbf{x})|^{2}\right]u_{r}(\mathbf{x})+U_{0}\phi^{2}(\mathbf{x})v_{r}(\mathbf{x}) &=& \epsilon_{r}u_{r}(\mathbf{x})\label{BdGTF1} \\
   \left[-\frac{\hbar^{2}}{2m}\nabla^{2}-\mu+V_{ext}(\mathbf{x})+2U_{0}|\phi(\mathbf{x})|^{2}\right]v_{r}(\mathbf{x})+U_{0}\phi^{*\,2}(\mathbf{x})u_{r}(\mathbf{x}) &=& -\epsilon_{r}v_{r}(\mathbf{x})\label{BdGTF2}.
\end{eqnarray}
Using the Thomas-Fermi result (\ref{ThomasFermi}) and the fact that $\phi$ is a real function we can approximate BdG equations as
\begin{eqnarray}
  \left[-\frac{\hbar^{2}}{2m}\nabla^{2}+U_{0}\phi^{2}(\mathbf{x})\right]u_{r}(\mathbf{x})+U_{0}\phi^{2}(\mathbf{x})v_{r}(\mathbf{x}) &=& \epsilon_{r}u_{r}(\mathbf{x}) \\
   \left[-\frac{\hbar^{2}}{2m}\nabla^{2}+U_{0}\phi^{2}(\mathbf{x})\right]v_{r}(\mathbf{x})+U_{0}\phi^{2}(\mathbf{x})u_{r}(\mathbf{x}) &=& -\epsilon_{r}v_{r}(\mathbf{x}).
\end{eqnarray}
From (\ref{GPSOL}) and (\ref{mutf}) we see that
\begin{equation}
  U_{0}\phi^{2}(\mathbf{x})=U_{0}C_{0}\frac{\sin k_{0}r}{k_{0}r}=\frac{GMm}{R_{0}}\frac{\sin k_{0}r}{k_{0}r}=|\mu_{TF}|\frac{\sin k_{0}r}{k_{0}r}.
\end{equation}
So for small $|\mu_{TF}|$ we can solve (\ref{BdGTF1}) and (\ref{BdGTF2}) perturbatively as in the case of a uniform condensate discussed at the end of Sec.\ref{BdGSection} and obtain
\begin{eqnarray}\label{BdGSOL}
  u_{\mathbf{k}} = e^{-i\mathbf{k}\cdot \mathbf{x}},\;\;\;\;\;v_{\mathbf{k}}=0,\;\;\;\;\;\epsilon_{\mathbf{k}}=\frac{\hbar^{2}k^{2}}{2m}.
\end{eqnarray}
Indeed, for ultralight dark matter $m\simeq 10^{-23} \textrm{eV} \simeq 1.6\times 10^{-42} \textrm{J}$. On the other hand taking the total galactic dark matter mass to be $M\simeq 10^{12}M_{\odot}\simeq 2\times 10^{42}\textrm{kg}$ and the radius of the DM halo to be $R_{0}\simeq 100\, \textrm{kpc}\simeq 3\times 10^{21}\textrm{m}$ (rough values for our own Milky Way) we get $|\mu_{TF}|= GMm/R_{0}\simeq 8\times 10^{-49}\textrm{J}$.

\subsection{Boundary Layer}\label{BLSec}
Consider the Fourier transform of the condensate wavefunction
\begin{equation}\label{fourier00}
  \widetilde{\phi}\left(\mathbf{k}\right)=\int d^{3}x\,\phi(\mathbf{x})e^{-i\mathbf{k}\cdot \mathbf{x}}.
\end{equation}
Using the TF result (\ref{GPSOL}), passing to spherical coordinates, and integrating over the angular variables we find
\begin{eqnarray}\label{fourier0}
\widetilde{\phi}\left(\mathbf{k}\right)=\frac{4\pi\sqrt{C_{0}}}{k\sqrt{k_{0}}}\int_{0}^{R_{0}}dr\,\sqrt{r\sin k_{0}r}\sin kr.
\end{eqnarray}
In Sec.\ref{DissipationSection} we will need to evaluate this integral for large $k=|\mathbf{k}|$. Although the integral is convergent the standard method of repeated integration by parts to derive its large $k$ asymptotics generates divergent terms
\begin{equation}
  \int_{0}^{R_{0}}dr \sqrt{r\sin k_{0}r}\sin kr=-\frac{1}{k}\left.\sqrt{r\sin k_{0}r}\cos kr\right|_{0}^{R_{0}}+\frac{1}{k^{2}}\left.\frac{\sin k_{0}r+r\cos k_{0}r}{2\sqrt{r\sin k_{0}r}}\right|_{0}^{R_{0}}+\ldots
\end{equation}
Here the first term vanishes while the second term diverges in the upper limit $R_{0}$. This divergence is somewhat similar to the one encountered in the study of trapped BEC of ultracold atoms \cite{pethick, lundth, fetter} where the gradient of $\phi_{TH}$ is singular at the boundary of the condensate and this in turn leads to a divergent result for the expectation value of the kinetic energy. The main problem is that the TF approximation, which neglects the kinetic term next to the nonlinear term, is no longer reliable near the boundary of the condensate where the latter vanishes. One therefore has a region called the boundary layer near the boundary of the condensate where the kinetic term is comparable to the nonlinear term. In the case of trapped BEC comparison with numerical calculations \cite{pethick,lundth} shows that the divergent integrals can be regularized by cutting them off at the boundary between the bulk (also called the exterior region, meaning exterior to the boundary layer) where the TF approximation is reliable and the boundary layer (also called the interior region) where TF theory does not work well. We will indeed follow that strategy to work out the asymptotic expansion of (\ref{fourier0}). However, we will postpone the study of the latter to Sec.\ref{Fourieranalysis} and in what follows examine first the structure of the boundary layer for the GPP system.

In order to estimate the size of the boundary layer let us divide both sides of (\ref{GPP1}) by $\mu$ and make the change of variable
\begin{equation}
  \mathbf{x}=\frac{R_{0}}{\pi}\boldsymbol{\xi},\;\;\;\;\;\;\phi(\mathbf{x})=\sqrt{C_{0}}\,\overline{\phi}(\boldsymbol{\xi}).
\end{equation}
Note that $\boldsymbol{\xi}$ and $\overline{\phi}$ are dimensionless variables. Thus (\ref{GPP1}) becomes
\begin{equation}
 \left[ \varepsilon \nabla^{2}_{\boldsymbol{\xi}}-1+\overline{V}_{\textrm{sc}}(\boldsymbol{\xi})+\overline{U}_{0}|\overline{\phi}(\boldsymbol{\xi})|^{2}\right]\overline{\phi}(\boldsymbol{\xi}),
\end{equation}
where
\begin{eqnarray}
\varepsilon&=&-\frac{\hbar^{2}\pi^{2}}{2mR_{0}^{2}\mu}=\frac{\hbar^{2}\pi^{2}}{2GMm^{2}R_{0}},\label{vareps}\\
 \overline{U}_{0} &=& \frac{U_{0}C_{0}}{\mu}\\
 \overline{V}_{\textrm{sc}}(\boldsymbol{\xi})  &=& \frac{1}{\mu}\,V_{\textrm{sc}}\left(\frac{R_{0}\boldsymbol{\xi}}{\pi}\right)=-\frac{Gm^{2}}{\mu}\left(\frac{R_{0}}{\pi}\right)^{2}C_{0}
 \int d^{3}\xi'\,\frac{|\overline{\phi}(\mathbf{\xi}')|^{2}}{|\mathbf{\xi}-\mathbf{\xi}'|}.
  \end{eqnarray}
Using the TF result (\ref{mutf}) we get
\begin{equation}
 \overline{U}_{0}=-1,\;\;\;\;\; \varepsilon=\frac{\hbar^{2}\pi^{2}}{2GMm^{2}R_{0}}
\end{equation}
Moreover, assuming the solution is spherically symmetric $\overline{\phi}(\boldsymbol{\xi})=\overline{\phi}(\xi)$, where $\xi=|\boldsymbol{\xi}|$, we obtain
\begin{equation}
  \varepsilon\left[\frac{d^{2}}{d\xi^{2}}+\frac{2}{\xi}\frac{d}{d\xi}\right]\overline{\phi}+\left[-1+\overline{V}_{\textrm{sc}}(\xi)+
  \overline{U}_{0}|\overline{\phi}(\xi)|^{2}\right]\overline{\phi}(\xi),
\end{equation}
and
\begin{equation}
 \overline{V}_{\textrm{sc}}(\boldsymbol{\xi})=\overline{V}_{\textrm{sc}}(\xi)=-\frac{Gm^{2}}{\mu}\left(\frac{R_{0}}{\pi}\right)^{2}\frac{2\pi C_{0}}{\xi}
 \int_{0}^{\infty}d\xi'\,\xi'(\xi+\xi'-|\xi-\xi'|)|\overline{\phi}(\xi')|^{2}.
\end{equation}
Upon the transformation
\begin{equation}
  \overline{\phi}(\xi)=\frac{f(\xi)}{\xi}
\end{equation}
we get
\begin{equation}
  \varepsilon\frac{d^{2}f}{d\xi^{2}}+\left[-1+\overline{V}_{\textrm{sc}}(\xi)+\overline{U}_{0}\frac{f^{2}(\xi)}{\xi^{2}}\right]f(\xi)=0.
\end{equation}
with
\begin{equation}
\overline{V}_{\textrm{sc}}(\xi)=-\frac{Gm^{2}}{\mu}\left(\frac{R_{0}}{\pi}\right)^{2}\frac{2\pi C_{0}}{\xi}
\int_{0}^{\infty}d\xi'\,\frac{\xi+\xi'-|\xi-\xi'|}{\xi'}|f(\xi')|^{2}.
\end{equation}
In order to estimate the size of the boundary layer we let $\xi=\pi-\delta \zeta$ with $\delta>0$, $\zeta>0$ and consider the region $\zeta>>1$, $\delta\zeta<<1$ \cite{fetter, bender}. Note that both $\delta$ and $\zeta$ are dimensionless and $\delta\zeta=0$ corresponds to the boundary of the condensate in the TF approximation. Let us make the transformation
\begin{equation}
  \psi(\zeta)=\delta^{-1/2} f(\pi-\delta \zeta).
\end{equation}
Then
\begin{equation}\label{finGP}
  \frac{\varepsilon}{\delta^{3/2}}\frac{d^{2}\psi}{d\zeta^{2}}+\left[-1+\delta^{1/2}\,\overline{V}_{\textrm{sc}}(\pi-\delta \zeta)+\delta^{3/2}\,\overline{U}_{0}\frac{\psi^{2}(\zeta)}{(\pi-\delta \zeta)^{2}}\right]\psi(\zeta).
\end{equation}
On the other hand from the TF result (\ref{GPSOL}) we get
\begin{equation}
  f(\pi-\delta \zeta)=\sqrt{(\pi-\delta\zeta)\sin(\pi-\delta\zeta)}.
\end{equation}
Thus for $|\delta\zeta|<<1$ we have
\begin{equation}
   f(\pi-\delta \zeta)\sim \sqrt{\pi\delta\zeta}
\end{equation}
and
\begin{equation}
  \psi(\zeta)=O(\delta^{0}).
\end{equation}
Thus in (\ref{finGP}) the kinetic term and the nonlinear term are balanced \cite{bender} for $\delta=O(\varepsilon^{1/3})$. So we may place the boundary between the bulk and the boundary layer at $r=R_{0}-\frac{R_{0}\delta}{\pi}$ with $\delta=O(\varepsilon^{1/3})$.

\section{Energy Dissipation in the Condensate}\label{DissipationSection}
In order to calculate energy dissipation we need the linear response function given in (\ref{response}). In terms of the condensate wavefunction and the fluctuation field the perturbation term $\mathcal{O}_{1}$ given in (\ref{po1}) is written as
\begin{eqnarray}\label{O1}
\mathcal{O}_{1}(\mathbf{x})&=& \phi(\mathbf{x})D\phi(\mathbf{x})+\mathcal{C}(\mathbf{x})+\eta^{\dag}(\mathbf{x})L\eta(\mathbf{x}),\\
  \mathcal{C}(\mathbf{x})&=&\phi(\mathbf{x})D\eta(\mathbf{x})+\eta^{\dag}(\mathbf{x})D\phi(\mathbf{x}),\\
  D&=&-\frac{\hbar^{2}}{2m}(-\partial_{x}^{2}+\partial_{y}^{2}),
\end{eqnarray}
Note that the first term in (\ref{O1}) is quadratic in the condensate wave-function and therefore $O(N_{0})$. However it does not contribute to energy dissipation since its commutator with the other terms of $\mathcal{O}_{1}(\mathbf{x})$ vanishes. On the other hand $\mathcal{C}(\mathbf{x})$, being linear in the condensate wave-function, is $O(\sqrt{N_{0}})$ and the term quadratic in the fluctuations is $O(1)$. This power counting is in accordance with the observations made in \cite{RS}. So in order to get the largest contribution $O(N_{0})$ to the linear response function (\ref{response}) we consider the commutator of $O(\sqrt{N_{0}})$ terms
\begin{eqnarray}\label{bees}
  [\mathcal{C}(\mathbf{x},t),\mathcal{C}(\mathbf{x}',t')] &=& \phi(\mathbf{x})\phi(\mathbf{x}')[D\eta(\mathbf{x},t),D'\eta(\mathbf{x}',t')]+\phi(\mathbf{x})D'\phi(\mathbf{x}')[D\eta(\mathbf{x},t),\eta^{\dag}(\mathbf{x}',t')]\nonumber \\
   &&+\phi(\mathbf{x}')D\phi(\mathbf{x})[\eta^{\dag}(\mathbf{x},t),D'\eta(\mathbf{x}',t')]+D\phi(\mathbf{x})D'\phi(\mathbf{x}')[\eta^{\dag}(\mathbf{x},t),\eta^{\dag}(\mathbf{x}',t')].\nonumber\\
\end{eqnarray}
The commutators of the fluctuations appearing in this expression are readily calculated from (\ref{etas0}),
\begin{eqnarray}
  [\eta(\mathbf{x},t),\eta^{\dag}(\mathbf{x}',t')] &=& \sum_{r} u_{r}(\mathbf{x})u^{*}_{r}(\mathbf{x}')e^{-i\omega_{r}(t-t')}-v^{*}_{r}(\mathbf{x})v_{r}(\mathbf{x}')e^{i\omega_{r}(t-t')},
 \end{eqnarray}
 \begin{eqnarray}
     [\eta^{\dag}(\mathbf{x},t),\eta(\mathbf{x}',t')]&=& \sum_{r} v_{r}(\mathbf{x})v^{*}_{r}(\mathbf{x}')e^{-i\omega_{r}(t-t')}-u^{*}_{r}(\mathbf{x})u_{r}(\mathbf{x}')e^{i\omega_{r}(t-t')},
 \end{eqnarray}
 \begin{eqnarray}
   [\eta^{\dag}(\mathbf{x},t),\eta^{\dag}(\mathbf{x}',t')] &=& \sum_{r} v_{r}(\mathbf{x})u^{*}_{r}(\mathbf{x}')e^{-i\omega_{r}(t-t')}-u^{*}_{r}(\mathbf{x})v_{r}(\mathbf{x}')e^{i\omega_{r}(t-t')},
 \end{eqnarray}
  \begin{eqnarray}
    [\eta(\mathbf{x},t),\eta(\mathbf{x}',t')] &=& \sum_{r} u_{r}(\mathbf{x})v^{*}_{r}(\mathbf{x}')e^{-i\omega_{r}(t-t')}-v^{*}_{r}(\mathbf{x})u_{r}(\mathbf{x}')e^{i\omega_{r}(t-t')}.
  \end{eqnarray}
  Using these in (\ref{bees}) we get
\begin{eqnarray}
  [\mathcal{C}(\mathbf{x},t),\mathcal{C}(\mathbf{x}',t')] &=& \sum_{r} \left\{e^{-i\omega_{r}(t-t')} \left[\phi(\mathbf{x})\phi(\mathbf{x}')Du_{r}(\mathbf{x})D'v_{r}^{*}(\mathbf{x}')+\phi(\mathbf{x})D'\phi(\mathbf{x}')Du_{r}(\mathbf{x})u_{r}^{*}(\mathbf{x}') \right.\right.\nonumber\\&&\left. +\phi(\mathbf{x}')D\phi(\mathbf{x})v_{r}(\mathbf{x})D'v_{r}^{*}(\mathbf{x}')+D\phi(\mathbf{x})D'\phi(\mathbf{x}')v_{r}(\mathbf{x})u_{r}^{*}(\mathbf{x}')\right]+\nonumber\\
  &&-e^{i\omega_{r}(t-t')}\left[\phi(\mathbf{x})\phi(\mathbf{x}')Dv_{r}^{*}(\mathbf{x})D'u_{r}(\mathbf{x}')+\phi(\mathbf{x})D'\phi(\mathbf{x}')Dv_{r}^{*}(\mathbf{x})v_{r}(\mathbf{x}')\right.\nonumber\\
&&\left.\left.+\phi(\mathbf{x}')D\phi(\mathbf{x})D'u_{r}(\mathbf{x}')u_{r}^{*}(\mathbf{x})+D\phi(\mathbf{x})D'\phi(\mathbf{x}')u_{r}^{*}(\mathbf{x})v_{r}(\mathbf{x}')\right]\right\}
\end{eqnarray}
Thus (\ref{response}) takes the form
\begin{eqnarray}\label{integrated1}
  &&\chi(\omega,\mathbf{x},\mathbf{x}') =\frac{i}{\hbar} \sum_{r} \left\{\left[i\mathcal{P}\frac{1}{\omega-\omega_{r}}+\pi\delta(\omega-\omega_{r})\right] \left[\phi(\mathbf{x})\phi(\mathbf{x}')Du_{r}(\mathbf{x})D'v_{r}^{*}(\mathbf{x}')+\right.\right.\nonumber\\&&\left.+\phi(\mathbf{x})D'\phi(\mathbf{x}')Du_{r}(\mathbf{x})u_{r}^{*}(\mathbf{x}') +\phi(\mathbf{x}')D\phi(\mathbf{x})v_{r}(\mathbf{x})D'v_{r}^{*}(\mathbf{x}')+D\phi(\mathbf{x})D'\phi(\mathbf{x}')v_{r}(\mathbf{x})u_{r}^{*}(\mathbf{x}')\right]+\nonumber\\
  &&+\left[i\mathcal{P}\frac{1}{-\omega-\omega_{r}}-\pi\delta(\omega+\omega_{r})\right]\left[\phi(\mathbf{x})\phi(\mathbf{x}')Dv_{r}^{*}(\mathbf{x})D'u_{r}(\mathbf{x}')+\phi(\mathbf{x})D'\phi(\mathbf{x}')Dv_{r}^{*}(\mathbf{x})v_{r}(\mathbf{x}')\right.\nonumber\\
&&\left.\left.+\phi(\mathbf{x}')D\phi(\mathbf{x})D'u_{r}(\mathbf{x}')u_{r}^{*}(\mathbf{x})+D\phi(\mathbf{x})D'\phi(\mathbf{x}')u_{r}^{*}(\mathbf{x})v_{r}(\mathbf{x}')\right]\right\}.\nonumber\\
\end{eqnarray}
Now we have
\begin{eqnarray}\label{integrated2}
 && \int d^{3}x d^{3}x'\, e^{-i\frac{\omega}{c}(z-z')}\chi(\omega,\mathbf{x},\mathbf{x}') = \frac{i}{\hbar}\int d^{3}x d^{3}x'\, e^{-i\frac{\omega}{c}(z-z')} D\phi(\mathbf{x})D'\phi(\mathbf{x}')\nonumber\\
&& \sum_{r} S_{r}(\mathbf{x},\mathbf{x}')\left\{\left[i\mathcal{P}\frac{1}{\omega-\omega_{r}}+\pi\delta(\omega-\omega_{r})\right]+
  \left[i\mathcal{P}\frac{1}{-\omega-\omega_{r}}-\pi\delta(\omega+\omega_{r})\right]\right\},
  \end{eqnarray}
where
\begin{equation}
  S_{r}(\mathbf{x},\mathbf{x}')=u_{r}(\mathbf{x})v_{r}^{*}(\mathbf{x}')+v_{r}(\mathbf{x})u_{r}^{*}(\mathbf{x}')+
  u_{r}(\mathbf{x})u_{r}^{*}(\mathbf{x}')+v_{r}(\mathbf{x})v_{r}^{*}(\mathbf{x}'),
\end{equation}
 and in arriving (\ref{integrated2}) from (\ref{integrated1}) we integrated by parts to make all $D$'s act on the mode functions. If we now make the substitution $\mathbf{x}\leftrightarrow\mathbf{x}'$ in the second integral of (\ref{integrated2}) we find
 \begin{eqnarray}
 &&\int d^{3}x d^{3}x'\, e^{-i\frac{\omega}{c}(z-z')}\chi(\omega,\mathbf{x},\mathbf{x}')=\frac{i}{\hbar}\int d^{3}x d^{3}x'\, e^{-i\frac{\omega}{c}(z-z')} D\phi(\mathbf{x})D'\phi(\mathbf{x}')\nonumber\\\label{integrated3}
  &&\sum_{r}S_{r}(\mathbf{x},\mathbf{x}')\left\{e^{-i\frac{\omega}{c}(z-z')}\left[i\mathcal{P}\frac{1}{\omega-\omega_{r}}+\pi\delta(\omega-\omega_{r})\right]+e^{i\frac{\omega}{c}(z-z')}
  \left[i\mathcal{P}\frac{1}{-\omega-\omega_{r}}-\pi\delta(\omega+\omega_{r})\right]\right\}.\nonumber\\
\end{eqnarray}
Taking the odd part of (\ref{integrated3}) in $\omega$ we get
\begin{eqnarray}
  \left[ \int d^{3}x d^{3}x'\, e^{-i\frac{\omega}{c}(z-z')}\chi(\omega,\mathbf{x},\mathbf{x}')\right]_{\textrm{odd}} =
  \frac{i\pi}{\hbar}\int d^{3}x d^{3}x'\,\phi(\mathbf{x})\phi(\mathbf{x}')DD' R(\omega,\mathbf{x},\mathbf{x}').
\end{eqnarray}
where we defined
\begin{equation}\label{Rdef}
  R(\omega,\mathbf{x},\mathbf{x}')=\sum_{r}\left\{S_{r}(\mathbf{x},\mathbf{x}')\left[e^{-i\frac{\omega}{c}(z-z')}\delta(\omega-\omega_{r})-e^{i\frac{\omega}{c}(z-z')}
  \delta(\omega+\omega_{r})\right]\right\}.
\end{equation}
Thus the general expression (\ref{ediss}) for the energy dissipation takes the form
\begin{eqnarray}
  \Delta E = \int_{0}^{\infty}d\omega\,\omega F(\omega)|\widetilde{h}(\omega)|^{2}.
\end{eqnarray}
where
\begin{equation}
  F(\omega)= \frac{1}{\hbar}\int d^{3}x d^{3}x'\,\phi(\mathbf{x})\phi(\mathbf{x}')DD'R(\omega,\mathbf{x},\mathbf{x}').
\end{equation}
On the other hand the energy of the GW is given by \cite{maggiore}
\begin{equation}
  E_{\textrm{gw}}=\frac{c^{2}}{16\pi G}\int d^{3}x\,\dot{h}^{2},
\end{equation}
which in frequency domain reads
\begin{equation}\label{egw}
  E_{\textrm{gw}}=\frac{Ac^{3}}{16\pi^{2} G}\int d\omega\,\omega^{2}|\widetilde{h}(\omega)|^{2}.
\end{equation}
Here $A$ is the cross sectional area of the condensate. Thus the fractional energy dissipation is given as
\begin{equation}\label{fraction}
  \frac{\Delta E}{E_{\textrm{gw}}}=\frac{16\pi G}{ R_{0}^{2}c^{3}}\frac{\int_{0}^{\infty}d\omega\,\omega F(\omega)|\widetilde{h}(\omega)|^{2}}{\int_{0}^{\infty} d\omega\,\omega^{2}|\widetilde{h}(\omega)|^{2}}.
\end{equation}
Assuming $|\widetilde{h}(\omega)|^{2}$ is sharply peaked at $\omega=\omega_{m}$ we get
\begin{equation}\label{fracapprox}
  \frac{\Delta E}{E_{\textrm{gw}}}\simeq \frac{16\pi G}{ R_{0}^{2}c^{3}}\frac{F(\omega_{m})}{\omega_{m}}.
\end{equation}
The above formulae are our main theoretical results giving the fractional change in the energy of GW passing through a non-relativistic, self-gravitating BEC with repulsive self-interactions. In order to proceed without the explicit solutions of the BdG equations we resort to the approximation (\ref{BdGSOL}) which leads to
\begin{equation}
   S_{\mathbf{k}}(\mathbf{x},\mathbf{x}')=e^{-i\mathbf{k}\cdot(\mathbf{x}-\mathbf{x}')},
\end{equation}
\begin{eqnarray}
  DD'R(\omega,\mathbf{x},\mathbf{x}') = \left(\frac{\hbar^{2}}{2m}\right)^{2} \int \frac{d^{3}k}{(2\pi)^{3}}\left\{\delta(\omega-\omega_{\mathbf{k}})(k_{x}^{2}-k_{y}^{2})^{2}
  e^{-i(\mathbf{k}+\frac{\omega}{c}\mathbf{\widehat{z}})\cdot \mathbf{x}}e^{i(\mathbf{k}+\frac{\omega}{c}\mathbf{\widehat{z}})\cdot \mathbf{x}'}\right\},
\end{eqnarray}
and
\begin{eqnarray}
  F(\omega) =\frac{1}{\hbar}\left(\frac{\hbar^{2}}{2m}\right)^{2}\int \frac{d^{3}k}{(2\pi)^{3}}\left\{\delta(\omega-\omega_{\mathbf{k}})(k_{x}^{2}-k_{y}^{2})^{2}
  \left|\widetilde{\phi}\left(\mathbf{k}+\frac{\omega}{c}\mathbf{\widehat{z}}\right)\right|^{2}\right\},
\end{eqnarray}
where $\widetilde{\phi}(\mathbf{k})$ is the Fourier transform (\ref{fourier00}) of the condensate wavefunction $\phi$.

Since $\phi=O(\sqrt{N_{0}})$ this term will give $O(N_{0})$ contribution to the energy dissipation. Note that this contribution would vanish in the case of a uniform condensate for which $D\phi=0$.

On the other hand passing to spherical coordinates in $\mathbf{k}$-space we have
\begin{equation}
  (k_{x}^{2}-k_{y}^{2})^{2}=k^{4}\sin^{4}\theta\,\cos^{2}2\varphi,\;\;\;\;\; \delta(\omega-\omega_{\mathbf{k}})=\frac{2m}{\hbar k}\delta\left(k-\sqrt{\frac{2m\omega}{\hbar}}\right),
\end{equation}
and we obtain
\begin{eqnarray}
  F(\omega) =\frac{\hbar^{2}}{2m}\frac{1}{(2\pi)^{3}}\int_{0}^{2\pi} d\varphi \cos^{2}2\varphi \int _{0}^{\pi} d\theta \sin^{5}\theta\int _{0}^{\infty}dk k^{5}\delta\left(k-\sqrt{\frac{2m\omega}{\hbar}}\right) \left|\widetilde{\phi}\left(\mathbf{k}+\frac{\omega}{c}\mathbf{\widehat{z}}\right)\right|^{2}.
\end{eqnarray}
Integrating over $k$ and $\varphi$ we find
\begin{equation}\label{fomega}
  F(\omega) =\frac{\hbar^{2}}{2m}\frac{\pi}{(2\pi)^{3}} \left(\frac{2m\omega}{\hbar}\right)^{5/2}\int _{0}^{\pi} d\theta \sin^{5}\theta\,\left|\widetilde{\phi}\left(\mathbf{k}+\frac{\omega}{c}\mathbf{\widehat{z}}\right)\right|^{2}_{k=\sqrt{\frac{2m\omega}{\hbar}}}.
\end{equation}

\subsection{Fourier Transform of the Condensate Wavefunction}\label{Fourieranalysis}
Let us now consider the large $k$ analysis of the Fourier transform $\widetilde{\phi}(\mathbf{k})$ of the condensate wavefunction. Making the change of variable $k_{0}r=\xi$ in (\ref{fourier0}) we get
\begin{eqnarray}\label{fourier}
  \widetilde{\phi}\left(\mathbf{k}\right)
=\frac{4\pi\sqrt{C_{0}}}{kk_{0}^{2}}\int_{0}^{\pi-\delta}d\xi\,\sqrt{\xi\sin \xi}\sin \frac{k}{k_{0}}\xi.
\end{eqnarray}
Here following the discussion of Sec.\ref{BLSec} we cut off the integral at the boundary between the bulk and the boundary layer.
Thus
\begin{equation}
  \left.\widetilde{\phi}\left(\mathbf{k}+\frac{\omega_{m}}{c}\mathbf{\widehat{z}}\right)\right|_{k=\sqrt{\frac{2m\omega_{m}}{\hbar}}}=\frac{4\pi\sqrt{C_{0}}}{k'k_{0}^{2}}
  \int_{0}^{\pi-\delta}d\xi\,\sqrt{\xi\sin \xi}\sin \frac{k'}{k_{0}}\xi.\label{four}
\end{equation}
where
\begin{equation}
  k'=\left| \mathbf{k}+\frac{\omega_{m}}{c}\mathbf{\widehat{z}}\right|=\sqrt{\frac{2m\omega_{m}}{\hbar}+\frac{\omega^{2}}{c^{2}}+2\sqrt{\frac{2m\omega_{m}}{\hbar}}\frac{\omega}{c}\cos\theta}.
\end{equation}
In the regime $k'>>k_{0}$, which will be justified shortly, the asymptotic behaviour of the $\widetilde{\phi}$ can be obtained by integrating (\ref{four}) by parts,
\begin{eqnarray}
 \left.\widetilde{\phi}\left(\mathbf{k}+\frac{\omega_{m}}{c}\mathbf{\widehat{z}}\right)\right|_{k=\sqrt\frac{2m\omega_{m}}{\hbar}} &=& -\frac{4\pi\sqrt{C_{0}}}{k^{'2}_{k_{0}}}\left[\left.\sqrt{\xi\sin\xi}\cos\frac{k'\xi}{k_{0}}\right|_{0}^{\delta-\pi}-
 \int_{0}^{\pi-\delta}d\xi\,\frac{\sin\xi+\xi\cos\xi}{2\sqrt{\xi\sin\xi}}\cos\frac{k'\xi}{k_{0}}\right] \nonumber\\
   &\sim &  -\frac{4\pi\sqrt{C_{0}}}{k^{'2}k_{0}} \sqrt{(\pi-\delta)\sin(\pi-\delta)}\cos\frac{k'\xi}{k_{0}}.
\end{eqnarray}
For $\delta<<\pi$ we can further approximate this as
\begin{equation}
 \left. \widetilde{\phi}\left(\mathbf{k}+\frac{\omega_{m}}{c}\mathbf{\widehat{z}}\right)\right|_{k=\sqrt\frac{2m\omega_{m}}{\hbar}}\simeq -\frac{4\pi\sqrt{C_{0}}}{k^{'2}k_{0}}\sqrt{\pi\delta}\cos\frac{k'\xi}{k_{0}}.
\end{equation}
So using (\ref{C0}) we arrive at
\begin{equation}
  \left|\widetilde{\phi}\left(\mathbf{k}+\frac{\omega_{m}}{c}\mathbf{\widehat{z}}\right)\right|^{2}_{\sqrt\frac{2m\omega_{m}}{\hbar}}\simeq \delta\frac{2\pi^{2}N_{0}}{k^{'4}R_{0}}\left(1+\cos\frac{2k'\xi}{k_{0}}\right).
\end{equation}
Finally for $k=\sqrt{2m\omega_{m}/\hbar}<< \omega_{m}/c$ we get
\begin{equation}
  \left|\widetilde{\phi}\left(\mathbf{k}+\frac{\omega_{m}}{c}\mathbf{\widehat{z}}\right)\right|^{2}_{\sqrt\frac{2m\omega_{m}}{\hbar}}\simeq \delta\frac{2\pi^{2}N_{0}}{R_{0}}\frac{c^{4}}{\omega_{m}^{4}}\left(1+\cos\frac{2\omega_{m}\xi}{ck_{0}}\right).
\end{equation}
Moreover assuming
\begin{equation}\label{ineq}
  \frac{\omega_{m}}{c}>>\sqrt{\frac{2m\omega_{m}}{\hbar}},
\end{equation}
we have
\begin{equation}\label{kprime}
  k'\simeq \omega/c
\end{equation}
 and
\begin{equation}\label{foursq}
  \left| \widetilde{\phi}\left(\mathbf{k}+\frac{\omega_{m}}{c}\mathbf{\widehat{z}}\right)\right|_{k=\sqrt{\frac{2m\omega_{m}}{\hbar}}}^{2}\simeq \frac{16\pi c^{6}\,C_{0}}{\omega_{m}^{6}}=
  \frac{4c^{6}N_{0}k_{0}^{3}}{\omega_{m}^{6}\pi}.
  \end{equation}
Now for GW frequency of $1\,\textrm{kHz}$ and dark matter particle mass $m\sim 10^{-23}\textrm{eV}/c^{2}$ we have
\begin{equation}
  \frac{\omega_{m}}{c}\simeq 2\times 10^{-5}\textrm{m}^{-1},\;\;\;\;\;\;\;\;\sqrt{\frac{2m\omega_{m}}{\hbar}}\simeq 9\times 10^{-32}\textrm{m}^{-3},
\end{equation}
and indeed (\ref{ineq}) holds. Furthermore, for $R_{0} \sim 100\, \textrm{kpc}\simeq 3\times 10^{21}\textrm{m}$  (rough value for our own Milky Way) from (\ref{R0}) we get $k_{0}\simeq 10^{-21}\,\textrm{m}^{-1}$ and therefore $k'>>k_{0}$.

\subsection{Fractional Energy Dissipation}
Substituting (\ref{foursq}) in (\ref{fomega}) we get
\begin{equation}
   F(\omega_{m})=\frac{1}{4}\hbar\omega_{m}\left(\frac{2m\omega_{m}}{\hbar}\right)^{3/2}\frac{N_{0}}{R_{0}}\delta\int_{0}^{\pi}d\theta\,\sin^{5}\theta
   \,\frac{1}{k^{'4}}\,\left(1+\cos\frac{2k'\pi}{k_{0}}\right).
\end{equation}
Since $k'>>k_{0}$ stationary phase approximation may be applied to the integral involving the cosine term. Since
\begin{equation}
  \frac{dk'}{d\theta}=0
\end{equation}
implies $\sin\theta=0$ the contribution of the cosine term will be negligible. On the other hand recalling (\ref{ineq}) and (\ref{kprime}), and using
$ \int_{0}^{\pi}d\theta\,\sin^{5}\theta=\frac{16}{15}$,
we arrive at
\begin{equation}
  F(\omega_{m})= \frac{4}{15}\hbar\omega_{m}\left(\frac{2m\omega_{m}}{\hbar}\right)^{3/2}\frac{N_{0}}{R_{0}}\left(\frac{\omega_{m}}{c}\right)^{-4}\delta
\end{equation}
Thus (\ref{fracapprox}) gives
\begin{equation}\label{fractional1}
  \frac{\Delta E}{E_{\textrm{gw}}}=\frac{256\pi^{2}}{45}n_{0}\,\ell_{p}^{2}\left(\frac{2m\omega_{m}}{\hbar}\right)^{3/2}\left(\frac{\omega_{m}}{c}\right)^{-4}\delta,
\end{equation}
where
\begin{equation}
  \ell_{p}=\sqrt{\frac{G\hbar}{c^{3}}}\simeq 1.6\times 10^{-35}\,m,
\end{equation}
is the Planck's length and
\begin{equation}
  n_{0}=\frac{N_{0}}{\frac{4}{3}\pi R_{0}^{3}}=\frac{M}{\frac{4}{3}\pi R_{0}^{3}m},
\end{equation}
is the mean number density of the condensate. Note that (\ref{fractional1}) can be expressed in terms of the wave-number $k_{m}=\omega_{m}/c$ of the GW and the wave-number $k_{q}=\sqrt{2m\omega_{m}/\hbar}$ of the quasi-particle of energy $\hbar\omega_{m}$ (as implied by the delta function in (\ref{fomega})) as
\begin{equation}\label{fractional2}
  \frac{\Delta E}{E_{\textrm{gw}}}=\frac{256\pi^{2}}{45}n_{0}\,\ell_{p}^{2}\frac{k_{q}^{3}}{k_{m}^{4}}\delta.
\end{equation}

\section{Conclusion}
Considering ultralight particles of mass $m=10^{-23}\,\textrm{eV}/c^{2}\simeq 1.6\times 10^{-42}\,\textrm{J}/c^{2}$ and taking the total condensate mass $M \simeq 10^{12}M_{\odot}\simeq 2 \times 10^{42}\,\textrm{kg}$, and galactic halo radius $R_{0}\simeq 100\,\textrm{kpc}$ (rough values for our own Milky Way) we have $N_{0}=M/m\simeq 10^{101}$ and $n_{0}\simeq 10^{36}\textrm{m}^{-3}$. For the peak GW frequency we take the value $1 \textrm{kHz}$ detected by LIGO and/or Virgo and find $\omega_{m}/c\simeq 2.1\times 10^{-5} \textrm{m}^{-1}$, and $(2m\omega_{m}/\hbar)^{3/2}\simeq 9.2\times 10^{-32}\textrm{m}^{-3} $.

On the other hand taking the cutoff $\delta=\varepsilon^{1/3}$, in accordance with the discussion of Sec.\ref{BLSec}, where $\varepsilon$ is given as in (\ref{vareps}) we get  $\delta\simeq 7\times 10^{-2}$ . Thus
\begin{equation}
  \frac{\Delta E}{E_{\textrm{gw}}}\simeq 4.8\times 10^{-46}.
\end{equation}

LIGO / Virgo signals come from distances of order of a billion ($10^9$) light-years, and the typical intergalactic distance is about a million light-years; meaning that the signal crosses of the order of a thousand galaxies on its way to us. Hence the expected energy loss fraction would be of the order of $10^{-51}$, and the amplitude correction due to energy absorption in presumed BEC DM halos of the order of $10^{-25}$-$10^{-26}$. Therefore, fortunately the standard siren estimations from these signals are safe from this effect; on the other hand, unfortunately the presumed BEC DM halos leave no discernible imprint on gravitational waves.

\section*{Acknowledgement} The authors thank O. T. Turgut for helpful comments.

\section*{Appendix: Linear response theory}

In this appendix we give a brief review of linear response theory and its application to the calculation of energy dissipation in a system subject to a time dependent perturbation \cite{pottier}. Consider a system with time independent Hamiltonian $H_{0}$. Let the system be in equilibrium and have the density matrix $\rho_{0}$ which commutes with $H_{0}$.  Then the statistical average of an observable $\mathcal{O}_{S}(t)$, (the subscript $S$ means the observable is in the Schr\"{o}dinger picture and we are considering the general case of a time dependent observable)
\begin{equation}
  \langle \mathcal{O}_{S}(t)\rangle=\textrm{Tr}\left\{\mathcal{O}_{S}(t)\rho_{0}\right\}.
\end{equation}
A time dependent perturbation $H'(t)$ will drive the system out of equilibrium and consequently the density matrix $\rho(t)$ will evolve in time according to the von Neumann equation
\begin{equation}\label{vN}
  \frac{d\rho(t)}{dt}=-\frac{i}{\hbar}[H_{0}+H'(t),\rho(t)].
\end{equation}
Treating $H'(t)$ as a perturbation the first order correction to $\rho(t)=\rho_{0}+\delta\rho(t)$ is given by the equation
\begin{equation}
  \frac{d\delta\rho(t)}{dt}=-\frac{i}{\hbar}[H_{0},\delta\rho(t)]-\frac{i}{\hbar}[H'(t),\rho_{0}],
\end{equation}
whose formal solution is
\begin{equation}\label{deltarho}
  \delta\rho(t)=e^{-\frac{i}{\hbar}H_{0}t}\left\{-\frac{i}{\hbar}\int_{-\infty}^{t}dt\,[H'_{H}(t), \rho_{0}]\right\}e^{\frac{i}{\hbar}H_{0}t}.
\end{equation}
Here and in what follows the subscript $H$ means the operator is in the Heisenberg picture defined with respect to the Hamiltonian $H_{0}$,
\begin{equation}
  H'_{H}(t)=e^{\frac{i}{\hbar}H_{0}t}H'(t)e^{-\frac{i}{\hbar}H_{0}t}.
\end{equation}
Thus the statistical average of $\mathcal{O}_{S}(t)$ is given by the Kubo formula:
\begin{eqnarray}\label{OM}
 \langle \mathcal{O}_{S}(t)\rangle_{\rho}  &=& \textrm{Tr}\left\{\mathcal{O}_{S}(t)\rho_{0}\right\}+\textrm{Tr}\left\{\mathcal{O}_{S}(t)\delta\rho(t)\right\}\nonumber\\
   &=& \textrm{Tr}\left\{\mathcal{O}_{H}(t)\rho_{0}\right\}-\frac{i}{\hbar}\int_{-\infty}^{t}dt'\,\textrm{Tr}\left\{\mathcal{O}_{H}(t)[H'_{H}(t'),\rho_{0}]\right\}\nonumber\\
   &=& \textrm{Tr}\left\{\mathcal{O}_{H}(t)\rho_{0}\right\}-\frac{i}{\hbar}\int_{-\infty}^{\infty}dt'\,\Theta(t-t')\textrm{Tr}\left\{[\mathcal{O}_{H}(t),H'_{H}(t')]\rho_{0}\right\}.
   \end{eqnarray}
Now consider the time derivative of the average energy of the perturbed system
\begin{eqnarray}
  \frac{dE}{dt}&=&\frac{d}{dt}\textrm{Tr}\left\{(H_{0}+H'(t))\rho(t)\right\}=\textrm{Tr}\left\{\frac{dH'(t)}{dt}\rho(t)+(H_{0}+H'(t))\frac{d\rho(t)}{dt}\right\}\nonumber\\
  &=&\textrm{Tr}\left\{\frac{dH'(t)}{dt}\rho(t)\right\}-\textrm{Tr}\left\{(H_{0}+H'(t))[H_{0}+H'(t),\rho(t)]\right\}=\textrm{Tr}\left\{\frac{dH'(t)}{dt}\rho(t)\right\}.
\end{eqnarray}
  Assuming the perturbation to be of the form $H'(t)=\sum_{a}J_{a}(t)\mathcal{O}_{a}$ where $\mathcal{O}_{a}$'s are time independent Schr\"{o}dinger picture operators, and $J_{a}(t)$'s are real valued functions of time we get
 \begin{eqnarray}
   \frac{dE}{dt} = \sum_{a} \dot{J}_{a}(t)\textrm{Tr}\left\{\mathcal{O}_{a}\rho(t)\right\}.
 \end{eqnarray}
By (\ref{OM}), to first order in perturbation theory this is given as
\begin{equation}\label{power}
  \frac{dE}{dt}=\sum_{a} \dot{J}_{a}(t) \textrm{Tr}\left\{\mathcal{O}_{aH}(t)\rho_{0}\right\}-
  \sum_{a,b}\int_{-\infty}^{\infty}dt'\,\dot{J}_{a}(t)\chi_{ab}(t-t')J_{b}(t').
\end{equation}
where
\begin{equation}
  \chi_{ab}(t-t')=\frac{i}{\hbar}\Theta(t-t')\textrm{Tr}\left\{[\mathcal{O}_{aH}(t),\mathcal{O}_{bH}(t')]\rho_{0}\right\}.
\end{equation}

If $J_{a}(t)\rightarrow 0$ as $|t|\rightarrow\infty$ then integration of (\ref{power}) gives
\begin{equation}
  \Delta E=-
  \sum_{a,b}\int_{-\infty}^{\infty}\int_{-\infty}^{\infty}dt\,dt'\,\dot{J}_{a}(t)\chi_{ab}(t-t')J_{b}(t').
\end{equation}
This can easily be generalized to field theory as
\begin{equation}
  \Delta E=-
  \sum_{a,b}\int \,dt\,d^{3}x\,dt'\,d^{3}x'\,\dot{J}_{a}(t,\mathbf{x})\chi_{ab}(t-t',\mathbf{x}-\mathbf{x'})J_{b}(t',\mathbf{x'}).
\end{equation}
with
\begin{equation}
  \chi_{ab}(t-t',x-x')=\frac{i}{\hbar}\Theta(t-t')\textrm{Tr}\left\{[\mathcal{O}_{aH}(t,\mathbf{x}),\mathcal{O}_{bH}(t',\mathbf{x'})]\rho_{0}\right\}.
\end{equation}

\end{document}